# Astronomical outreach and education in marginalised and indigenous communities: astronomy as a tool for social development

Arianna Cortesi, Claudia Mignone, Alan Alves Brito, Gracy Moreira, Claudia Magnani, Nilson Moreira, Guilherme Moreira, Roberto Romeiro, Andrè Victor, Maria Clara Herdinger Lourenço, Gabriela Rufino Travasso, Ana Gomes, Lorena Anástacio

Observatorio do Valongo, Universidade Federal do Rio de Janeiro, Rio de Janeiro, Brazil; Istituto Nazionale di Astrofisica, Rome, Italy; Institute of Physics, Federal University of Rio Grande do Sul, Porto Alegre, Brazil; Organização Cultural Remanescentes de Tia Ciata, Rio de Janeiro, Brazil; Independent Researcher; Organização Cultural Remanescentes de Tia Ciata, Rio de Janeiro, Brazil; Organização Cultural Remanescentes de Tia Ciata, Rio de Janeiro, Brazil; Federal University of Minas Gerais, Belo Horizonte, Brazil; Universidade do Estado do Rio de Janeiro Centro de Tecnologia e Ciências Escola Superior de Desenho Industrial, Rio de Janeiro, Brazil; Observatorio do Valongo, Universidade Federal do Rio de Janeiro, Rio de Janeiro, Brazil; Observatorio do Valongo, Universidade Federal do Rio de Janeiro, Rio de Janeiro, Brazil; Federal University of Minas Gerais, Belo Horizonte, Brazil; Federal University of Minas Gerais, Belo Horizonte, Brazil.

**Abstract.** The way we look at the sky is connected to the cosmological paradigm embraced by the society we live in. On the other hand, several astronomical concepts reinforce the idea of a common humanity. Yet, scientific outreach is frequenty reaching out only to a specific part of the world population, often excluding people living in extreme social vulnerability, victims of violence and prejudice, fighting for their lives and for the right of living according to their traditions. We present two outreach projects, developed in Brazil, funded by the Office of Astronomy for Development (OAD) of the International Astronomical Union (IAU), i.e. "Under Other Skies" & "OruMbya", which tackle the importance of ethno-astronomy, and the collaboration with leaders and cultural agents of marginalised communities. We also describe an educational project born in the favela of Cantagalo Pavão Pavãozinho (PPG), in Rio de Janeiro, during the COVID19 pandemic, which started a collaboration with local educators and artists to offer classes of astronomy and English language to children in the favela.

**Keywords.** Astronomical outreach; astronomy for development; Social Development Goals

## 1. Introduction

The sky offers humanity a window to study the origin and evolution of the universe, from the Big Bang till the formation of the solar system and planet Earth. Beneath this sky, the contemporary socio-economical structure developed during centuries and millennia, leading to the highly unequal distribution of resources, and life quality and expectancy we observe nowadays, see Figure 1 for a global comparison between light pollution levels, population size, and indexes of violence. In this context, astronomy can be used as a unifying tool for the development and promotion of peace among peoples, fighting inequality and supporting cultural integration, as well as the restoration of dignity to 'other' conceptions of the world against the prejudices connected to a unimodal vision of history. For example, astronomy based projects have been supporting a multicultural approach to education, developing the synergy between scientific and cultural perspectives to produce an 'experimental togetherness' (Keller,

Catherine & Daniell, Anne (eds.) 2002). Moreover, pioneers works have been promoting environment preservation, as by contributing to the world-wide knowledge of Brazilian indigenous lands, which are target of exploitation and deforestation.

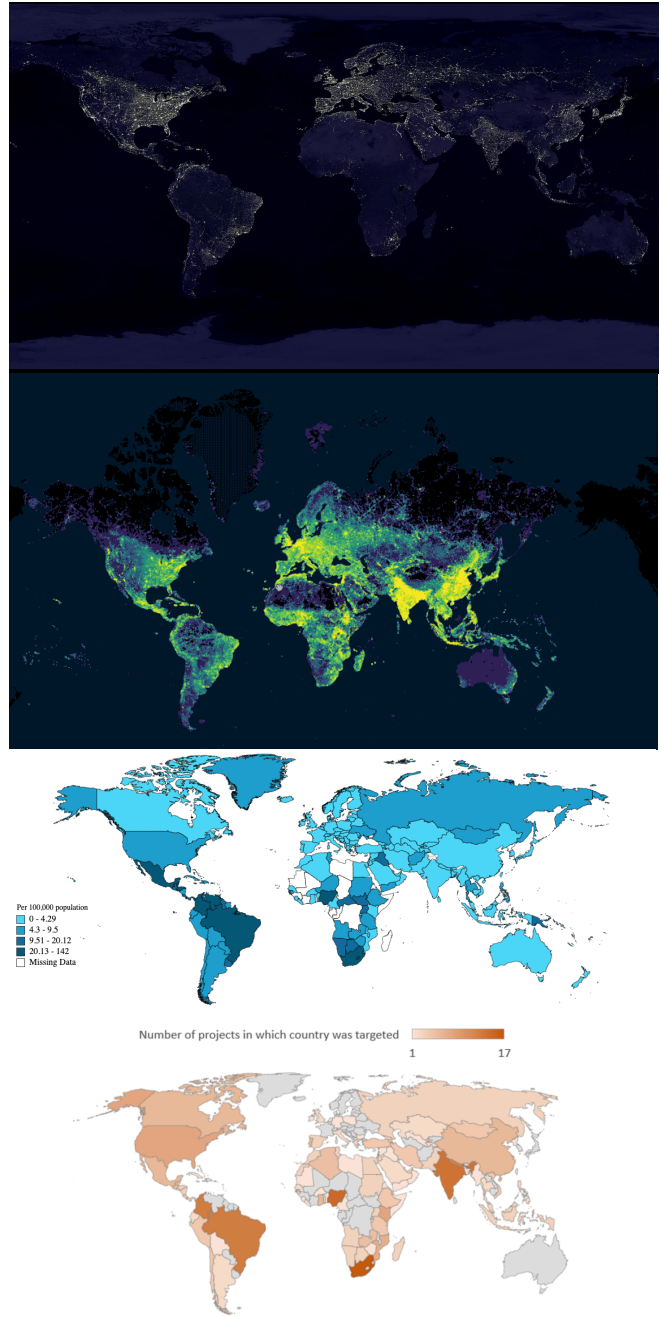

**Figure 1.** Top: A global view of Earth at night, compiled from over 400 satellite images (Image Credit: NASA/NOAA). Second row: Global population density, reference Period: March 11, 2020-November 01, 2023 (Image Credit https://images.app.goo.gl/1T2a9nyfvMSnJAiT7). Third row: Intentional homicides, per 100,000 population (Image Credit: World Bank SDGs database). Bottom: Number of OAD projects in different countries (Image Credit: https://www.astro4dev.org/projects-impact/)

In particular, the Office of Astronomy for Development† is a joint project of the International Astronomical Union and the South African National Research Foundation with the support of the Department of Science and Technology, which aims to pursue social development through funding and coordinating projects based on astronomy. Outcomes and deliverables of these projects, such as increasing in the literacy levels, or knowledge of astronomy, and production of novel curriculum resources, or teacher training material, encompass different perspectives. From a teaching/school perspective they support the training of teachers in the field of intercultural education, promoting the principle of inextricable connection between different areas of teaching, research and education of undergraduate students. From a community point of view, they promote a peaceful and welcoming environment in specific areas (such as marginalised and vulnerable communities), often suffering a historical profile of material and symbolic violence, leveling cultural and social differences, by increasing knowledge of cultural astronomy heritage, as a key aspect of the culture and daily life. From an academic perspective, debating, in an interdisciplinary way, scientific and cultural concepts on the visible sky, linking past and present, and the connections between the sky (cultural identity) and the Earth (territory, diaspora), these projects strengthen the relations between the University and School sectors, as well as with the Society at large. Moreover, given the world wide connections provided by the IAU/OAD, these innovative projects promote intercultural dialogues between knowledge among continents.

One of the countries hosting a high number of OADs projects is Brazil, see Figure 1, bottom panel, which due to its colonial past, suffered some of the most tragic episodes of human history, such as the transatlantic trade of enslaved people from Africa, leading to one of the largest African diaspora, the massacre of indigenous populations and a violent deforestation. Scars of this past still bleed in Brazilian present and are clearly visible in the inequality of the society, the racism and violence toward Black† and indigenous population, and the continuous devastation of the Amazon forest. Moreover, while astronomical outreach is becoming a fundamental step toward the integration of scientific concepts in everyday life and a point of connection between citizens and scientists (Simpson, Robert, Kevin R. Page, and David De Roure 2014), it is often restricted to the wealthy part of society (DAWSON, E. 2014). Recently, several projects started being created together with people that have long been kept out of the formal scientific process, fostering marginalised societies as examples where we can understand, learn and make new science (Benyei, P. 2023). In this work we present three projects developed in Brazil within the OAD/IAU framework and financial support, co-created in collaboration with marginalised communities, tackling several SDG goals such as quality education (SDG-4), reduced inequality within and among countries (SDG-10), promoting sustained, inclusive and sustainable economic growth, full and productive employment, and "decent work for all" (SDG-8), peace, justice, and strong institutions (SDG-17), "partnership for the Goals" (SDG-18), and gender equality (SDG-5)‡.

## 2. Under other skies: dialogues of different cosmological paradigms

This project aims to bridge different visions of the Universe we live in by collecting, in a webpage, indigenous knowledge, as narrated in traditional chants (as sung by pajes - spiritual leaders of the villages), carried out by the Other Skies team in an indigenous village in Minas Gerais, Brazil. The webpage also contains drawings of the stories (chants) created by indigenous professors, during a workshop on illustrations (see Figure 2 as an example), developed in the context of the course Educação Indígena "Saberes Indígenas nas Escolas" at the

---

† https://www.iau.org/development/oad/
† https://flacso.org.br/files/2016/08/Mapa2016_armas_web-1.pdf
‡ See :https://sdgs.un.org/goals for more details on the SDG goals.

Federal University of Minas Gerais (UFMG). The project is producing cultural and astronomical materials for indigenous and non-indigenous schools, and the participants are indigenous professors of Maxakali and other ethnic origins.

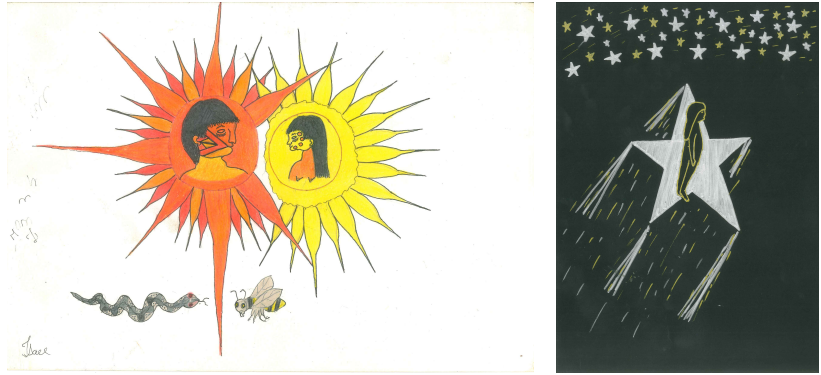

**Figure 2.** Left: A drawing representing the chant 'Brother Sun, brother moon'. Right: A drawing representing the chant 'The star women'.

The Other Skies team is composed of two indigenous professors and chapes (leaders) of the Village Aldeia Verde (BR), four anthropologists from the department of education of UMG (BR), four astronomers (BR and UK), one educator and one creative practitioner in community engagement and citizen science (UK).

The project was mostly developed during the COVID19 emergency, and the quarantine regime in the Brazilian territory. In the specific case of indigenous Villages in Brazil, FUNAI (Fundação nacional do Índio) with a 'Boletim de Serviço do 17 de março de 2020' resolved that: 'The granting of new authorisations to enter indigenous lands is suspended, with the exception of those necessary for the continuity of the provision of essential services to communities, as assessed by the competent authority of the Regional Coordination - CR'†. Due to complex conditions created in the first months of the isolation dictated by the pandemic, our collaborator professors of the village Aldeia Verde, together with dozens of families of the village, moved to a new village, with no infrastructure. As a consequence, to carry out the project as a virtual collaboration, it was necessary to install an internet connection in the new village and to provide instruments of communication. We applied for - and obtained - an extra-grant from IAU/OAD and we started looking for internet providers in this remote territory. Normally internet radio connection is used in this region of Brazil, but the existence of a gigantic stone between the transmitter and the village was rendering the use of the radio impossible. While we were looking for an alternative possibility, such as satellite internet (HUGHESNET) or bringing cable internet to the new village, the possibility arose that the indigenous community would move toward another land, due to the presence of a dam near the village that might compromise their safety and peaceful living. The internet installation was then paused, till the land problem was resolved. Even so, through the process of moving during the pandemic, several materials have been produced as a short movie shot by our collaborator Sueli Maxakali and edited by André Victor and Cristiano Araújo, which aimed to record an encounter of pajes in the village to discuss the COVID19 emergency‡ Our team also gave support to the preparation of English subtitles.

† Fica suspensa a concessão de novas autorizações de entrada nas terras indígenas, à exceção das necessárias à continuidade da prestação de serviços essenciais às comunidades, conforme avaliação pela autoridade competente da Coordenação Regional - CR.

‡ https://drive.google.com/file/d/19jWgObkVZNvFdCUN5-H_IQzsPGVclvsp/view?usp=sharing.

In April 2021, a new land was found, named Aldeia Nova, where the people could permanently reside, with running water and enough land to live, according to their way of life. We successfully installed two antennas and a stable connection for the entire community. Figure 3, left panel, s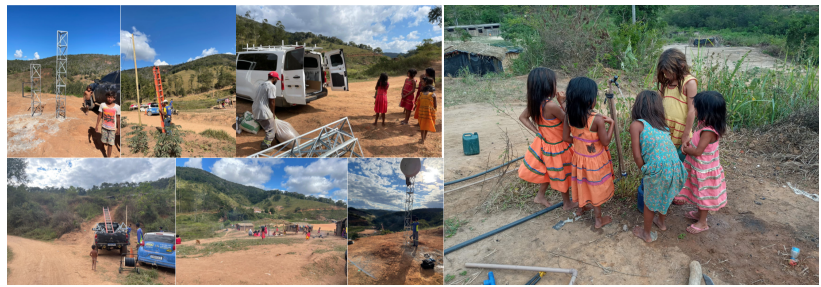ic                   as. The connection was successfully working an                                    discussion of the Undergraduate thesis of Sueli                                   `ps://www.youtube.com/watch?v=0fsc5M3af`          is            al online courses and meetings with the authori                      u       mple, the organisation of the exposition of th                      a       nale Art Exposition in São Paulo - September

**Figure 3.** Left: Two 10 meters high antennas for internet connection being built in the village Aldeia Nova, funded by the IAU/OAD extra grant. Right: Children at the artesian wall in The Aldeia Escola Floresta.

On the 17th of September 2021 Dr. Roberto Romero visited the village Aldeia Nova in order to coordinate the collection of material (drawings and chants) in collaboration with four young professors of the village. During the visit of Dr. Romero, the community moved to a new Aldeia, where they are building the Village Forest School†. With the OAD grant we were able to support the construction of an artesian well, lead by the Theofilo Otoni Caritas. Figure 3, right panel, pictures some children of the Aldeia using the well.

The result of the project, apart for being collected in the website, was published in a paper for an on-line blog, Big Data from the South Research Initiative (BigDataSur), 'a space for theoretical and empirical exchange on the challenges of datafication and massive data collection as they unfold in the plurality of South(s) inhabiting our increasingly complex world'. Specifically, we contributed to the section of the blog called 'COVID-19 from the margins' dedicated to forgotten and silenced voices of marginalized and disadvantaged communities and populations. Our contribution, entitled 'Under Other Skies: Astronomy as a Tool to Face COVID-19-Induced Isolation in the Indigenous Village of Aldeia Verde, Brazil', reports the story of the project through the pandemic and importance of the support of IAU/OAD, as well as the ever-present sustained support of the department of education of UFMG, especially Roberto Romero, Ana Maria Gomes and Paola Gobetti. Our contribution was also published in the book 'COVID-19 from the margin, pandemic invisibilities, policies, and resistance in the datafied society', which resulted by collecting all the stories from the blog.

## 3. OruMbya - Astronomy as fuel of life: the resilience of stars in Yoruba, Afro-Brazilian and Indigenous Cosmogony

OruMbya (Orum, sky in Yorubá, and Mbya, a Brazilian Guarani ethnicity) is a project to celebrate Astronomy as the fuel of life, in which the stories of the stars are preserved in the resilience of people from three different continents (South America, Europe and Africa) and shared over months, through scientific and cultural activities focused on the dissemination of knowledge, promotion of social inclusion and sustainable development in the context of PLOAD (Portuguese Language Office of Astronomy for Development). In its first realisation, lead by Doctor Honoris Causa Gracy Mary Moreira, the hosts were the cultural centre "Remanescentes da Tia Ciata", whose aim is the defence and preservation of afro-brazilian memories in Rio, and the centenary Observatory of Valongo (OV), one of the oldest institutions of Astronomy in Brazil. The two institutions are located in Morro da Conceição, an iconic place of resistance and reaffirmation of black identity, one of the most socially vulnerable urban regions in Rio. UFRGS (The Federal University of Rio Grande do Sul) is the third partner of the project.

† https://aldeiaescolafloresta.org/

The project was developed by taking into account three possible degrees of impact of COVID-19 on social distancing: none, moderate and total. The rules of social distancing changed in Brazil during the nine months of development of the project. Specifically, we held all of the webinars online, see Figure 4, apart from the last seminar that was held in mixed mode (i.e., most of the participants of the projects joined together at the observatory, while the event was streamed in the usual platform).

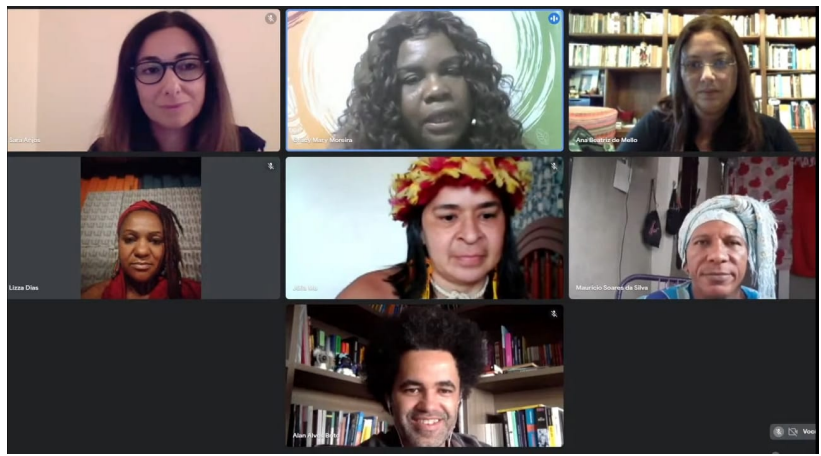

**Figure 4.** A Screenshot of one of the webinars realised during the OruMbya project, picturing the different speakers present, representing different sectors of knowledge

The 5 webinars (with teachers, students, Indigenous, Quilombolas and Favelas representatives and astronomers) connected people of very different cultural backgrounds, who share common goals of a just, peaceful and green future, telling stories and sharing knowledge about culture and the sky, creating new multicultural collaborations and alliances. We had more than 17 speakers from different segments (in even number) participating in the webinars and more than 1000 total views. Two hundreds and two persons filled the online questionnaire and sent us a brief summary of their thoughts on the project and the subjects discussed, actively contributing to the project.

At the observatory we held a workshop of plantation of species from the indigenous and Afro-Brazilian culture in collaboration with 15 children from the territory (the number is restricted to meet the measures of social distancing due to the pandemic) and their teacher, coming from a nearby favela (Favela da Providencia - ONG Impacto das Cores). Moreover, we offered 2 online extension courses of 2hrs each for teachers, creating new literature and enriching the possible way of study offered by the observatory, the university and the ORTC, creating new knowledge. Finally, we produced two videos to promote the project and the territory, changing the vision of the little Africa region of Rio de Janeiro, fighting for a renaissance of the area and preservation of its culture and monuments, as the world heritage Cais do Valongo, which risks losing this status. The pilot project sprouted in several subsequent projects as 'Orumbya -Mulheres do mundo sócio-cultural-tecnológico' (PI: Gracy Mary Moreira), a project aimed at girls aged between 14 and 20, interested in science. Designed in collaboration between Casa da Tia Ciata, the Valongo Observatory (OV/UFRJ), the Federal University of Rio Grande do Sul and the State University of Rio de Janeiro (UERJ), the project was awarded the 2nd edition of the "STEM Girls: training future scientists"† project, promoted by the British Council Brazil and the Carlos Chagas Foundation, see Figure 5 for an example of the flyer promoting the classes, presenting the two professors, from different fields of knowledge. In 2023, we are developing the project 'OruMbya – a library of silenced voices‡' (PI: Alan Alves Brito), whose main goal is to strengthen primary and secondary education in science, through equity and equal opportunities for children and young people living in extreme social vulnerability.

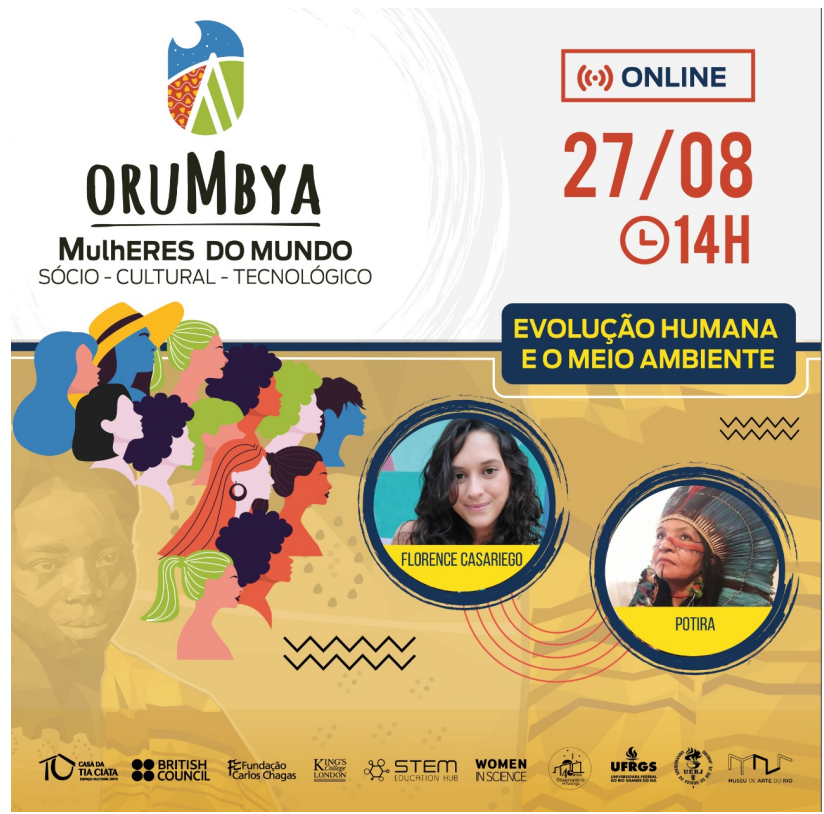

**Figure 5.** A flyer advertising the first class of the course 'Orumbya Mulheres' (https://ov.ufrj.br/orumbya-mulheres-do-mundo-socio-cultural-tecnologico/).

## 4. Closer to the Sky

This project aims at co-producing scientific knowledge in collaboration between astronomers and artists/educators living in the favela of Cantagalo Pavão e Pavãosinho (PPG), Rio de Janeiro (Brazil), for children, teenagers and young adults of the community. It is the result of the close collaboration between The PPG Astronomical Club, whose logo, see Figure

---

† Science, Technology, Engineering and Mathematics
‡ https://www.astro4dev.org/orumbya-library-of-silenced-voices/

6, pictures an artistic view of the PPG community in the night sky, and the social project 'Ninho das aguias'†, where classes and night sky observations are held.

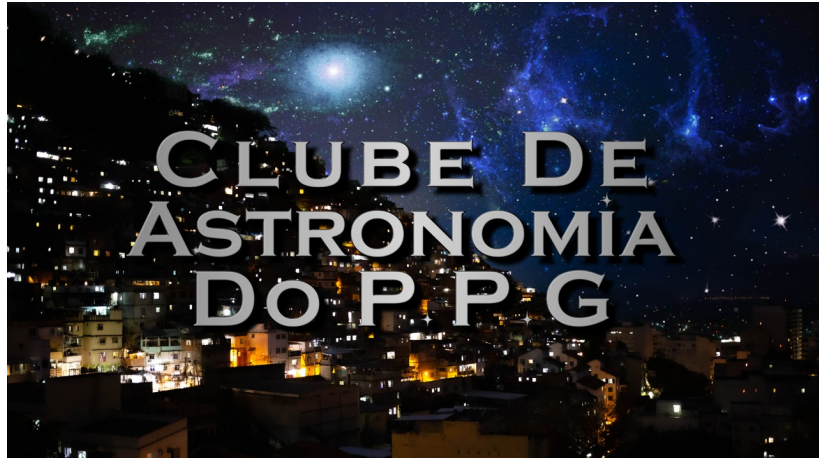

Figure 6. Artistic representation of the night sky in the favela do Cantagalo Pavão e Pavãozinho (Rio de Janeiro, Brazil), used as logo of the astronomical club.

Favelas in Rio de Janeiro state are often located on hills, which served as refuge for enslaved people - after the abolition of slavery (1888) - and immigrants, who worked for downtown citizens. The Law of lands (1850) prevented unoccupied lands to be owned through labour and provided government subsidies for the arrival of foreign settlers to be hired in the country, further devaluing the work of black men and women. As a result, favelas today are mostly composed of a black population, surviving decades of persecutions and low income while defending, preserving and creating a unique culture rooted in African origins that reverberates into music and the arts. Children in favelas, due to social and economical inequality and racial discrimination, have fewer opportunity for personal development and professional realization. They attend public schools where in 2021, according to SAEB‡, students do not reach a satisfactory level in Portuguese language (69%) and mathematics (95%). They hardly have access to after-school courses and do not tend to see themselves represented in the academic community. This produces a disadvantage in access to higher education and consequently in opportunities for satisfactory employment. Offering extracurricular courses and cultural experiences to students in the PPG, we wish to enrich their school curriculum and strengthen the chance that they stay in education after secondary school. A key element of the courses is providing positive role models of scientists from afro-descendant backgrounds, reinforced by the presence of local artists and educators, thus endorsing their role within the academic community. Specifically, we aim at using the arts and the act of embodiment (Radford, Luis, 2011) as a tool to improve the learning experience (see Figure 7 for an example of a workshop on the stellar life cycle).

† https://www.instagram.com/ninhodasaguiasppg/
‡ https://www.gov.br/inep/pt-br/areas-de-atuacao/avaliacao-e-exames-educacionais/saeb/resultados

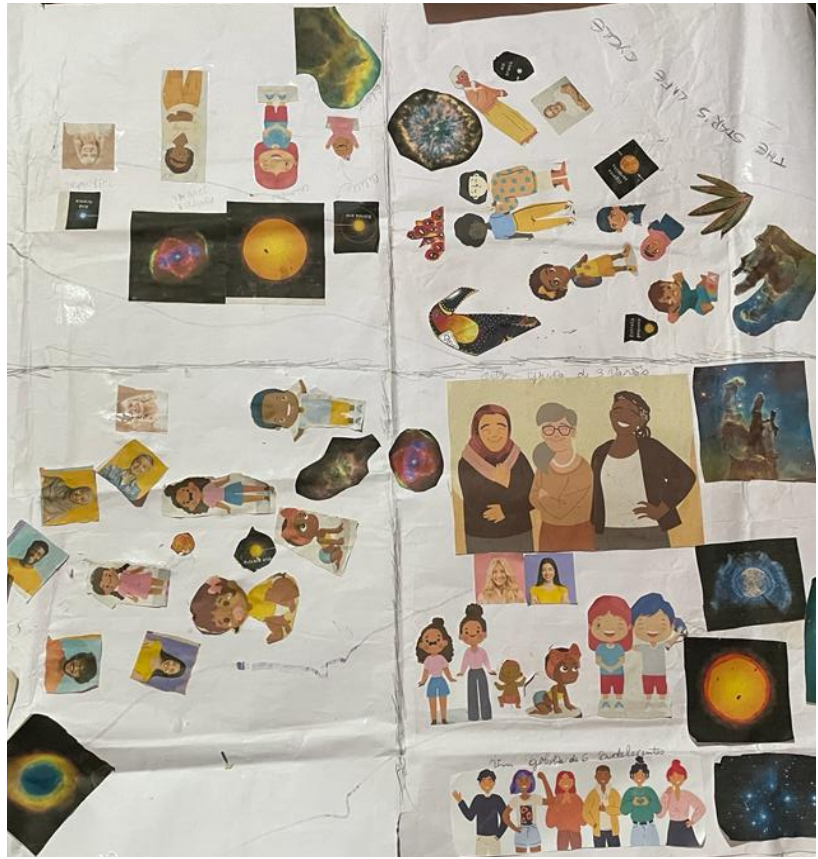

**Figure 7.** Example of a cooperative student work aimed at comparing the phases of life of stars and of humans.

The project also creates work opportunities for local artists and educators, who will offer workshops rooted in favela culture, while at the same time creating novel, de-colonial course ware based on contextualised science, i.e, materials that use the context of marginalised societies as examples where we can understand, learn and make science. The material developed within the project is shared as 'open educational resources'. In the next year, five students will be trained as astroguides, in collaboration with Astronera, a project committed to the scientifically accurate and innovative dissemination of astronomy, it was established by Shweta Kulkarni (Bangalore, India) †. Their training will benefit from visits and courses at the local Planetarium and Museum of Astronomy.

## 5. Discussion and conclusion

Stengers (2002) underlined how multicultural approaches to education are fundamental for pursuing sustainable development. The projects discussed here propose a dialogue between different visions of the sky, opposing the idea that one scientific paradigm is superior to another, following the steps of many fundamental works, see, for example, Hugh-Jones, S. (1979); Walmir Cardoso's (2007); Oscar T. Matsuura (2013); Afonso, G. B. (2014); Alan Alves-Brito et. al (2019). The team of researchers involved in all the projects is heterogeneous in gender, ethnicity and field of study (comprising astronomers, anthropologists, school teachers, indigenous school teachers and researchers, artists and museum managers), generating a fair and respectful approach to the subject, with participants carrying out the activities in full collaboration.

† https://www.astronera.org/

These projects are examples of the heterogeneity of astronomical knowledge. Through their intercultural approach, they promote an understanding of the different astronomical paradigms, and push back against a superficial approach to science and social biodiversity†. Introducing those concepts in official centres of knowledge and publishing the material in university-based web pages, proceedings and various media of communication is a way of fighting stereotypes and reducing discrimination, contributing to a more informed understanding of indigenous people and marginalised communities and their way of life, increasing the inhabitants' political power. In addition, these projects raise awareness of exploitation and deforestation on national and global scales. The activities are included in the university programs, when possible, such as for example at FIEI (Intercultural Education for Indigenous Teachers) of the UFMG, opening new lines of academic study, and initiating a collaboration between the departments of astronomy and education.

The main challenges of these projects are (i) to promote, through Astronomy in Cultures, the intercultural dialogue between scientists and researchers from other fields and the general public; and (ii) to promote a new inclusive scientific literacy focused on the initial and continuing teaching training process in Brazil, where the contribution of underrepresented people (women, black and indigenous) are recognised and valued in the first place.

On the one hand, the objectives of social development, pursued by each institution independently, are supported and expanded through the collaboration between them, by promoting astronomy and science as part of the culture (race, gender, class, education) in Morro da Conceição, an iconic place of resistance and reaffirmation of black identity, one of the most socially vulnerable urban regions in Rio, in the favela of Cantagalo Pavão Pavãozinho (Rio de Janeiro, Bazil) and in the indigenous Florest School Village (Estado de Minas Gerais, Brazil). On the other hand, the virtual activities give the projects more visibility and allow us to invest not only in the training of teachers around Brazil, but also to expand our cultural/educational projects (black power, memory, cultural patrimony) to other lusophone countries (Portugal, and countries in Africa and Asia).

Focusing on the creation of a novel scientific literacy, teacher training, and the intercontinental cultural astronomy exchange, working with positive models of women and LGBT+ people in Astronomy, making their work visible, and helping people from different social backgrounds to develop their skills, opening access to the observatory to the afro-brasilian and indigenous communities, explicitly recognising the scientific and social relevance of their inherited knowledge, promoting the sustainable economic growth of ONGs, globally advertising the existence of these marginalised communities, creating a path towards the community revival of vulnerable areas of the city, are some of the main points tackled by these projects.

Astronomy is a unique science for its potential to inspire and fascinate people around the world. Recent studies, including an IAU OAD flagship project (Vertue, D. G. 2022), have been investigating the restorative power of stargazing and contemplating the universe to people's mental well-being. These initiatives promote local artists and educators and create a local network of cultural projects inspired by and making use of astronomical knowledge, putting them in contact with universities and research centres in the city of Rio and beyond, opening the possibility for future collaborations and employment. They support and foster local business, and create a safe knowledge space where youth feel welcome and have access to both material and immaterial resources, encouraging them to continue their studies in higher education, and achieve better paid jobs.

Measuring the impact of astronomy for development projects poses several challenges in terms of time, cost, accurate data and cooperation (Chapman, S. 2015). This is especially relevant when working with marginalised communities in fragile contexts, in order to protect these communities and their members. For the Orumbya projects, the participants, from very different segments, were encouraged to share narratives, videos and pictures, to capture their perception of the world. The webinars had more than a 1000 total views and the online courses were attended by 30 school professors and university stuff. Finally, we had the presencial participation of 15 children from a nearby favela, as well as their teacher, during the plantation of the allotment. Under Other Skies project is still ongoing. The stories of the Maxakali cosmogony as well as their culture and fight for land preservation have been presented in several conferences, as at the world's largest conference on astronomy communication: Communicating Astronomy with the Public 2022 (CAP 2022). Moreover, the project had a concrete impact in the day life of the indigenous village, with the construction of the two antennas for internet provision and the implementation of an artesian well. As for the more recent project, Closer to the Sky, a broader evaluation plan is envisioned, monitoring the wellbeing and self-esteem of children with support from medical and public anthropology, as well as course attendance of children and their long-term schooling activities, and the activities of the newly trained astroguides as they search for new work opportunities. The education interventions will be monitored with support from science education experts by evaluating the learning experience of children and teenagers pre- and post-course through questionnaires for self and peer assessment to empower learners and promote collaboration. We also envision conducting focus groups at the end with adults who have taken part in the project as collaborators, as well as families of children who have attended the course, to assess their perceptions of the project and of its impact on the children and the community.

---

† We refer the readers to the projects listed in the OAD/IAU webpage (https://www.astro4dev.org/projects-search/) for an extended and global perspective.

By rediscovering cultural astronomy, for example traditional constellations, telling their stories - and the science underneath - in many different artistic formats, as well as empowering people often discriminated against on the basis of their colour and social origins, these projects centre them as the person of knowledge, promoting cultural dissemination, as well as opening the doors of Universities, Museums and Planetariums to people who are often excluded due to prejudice, cost and content perceived as "not for us" (DAWSON, E. 2014). These projects change who makes science, promoting self-representation among marginalised groups. Only co-creating knowledge by and for people that have long been kept out of the formal scientific process, will research truly be open.

We acknowledge funding from FAPERJ grant E-26/200.607 e 210.371/2022(270993) and of the OAD/IAU projects "Under Other Skies" and "OruMbya". We thank Ramasamy Venugopal for the kind support.